\documentclass[preprint,proceedings]{rmaa}

\suppressfulladdresses % by popular request

\usepackage{paralist}

\usepackage{psfrag,color}

\SetYear{2007}
\SetConfTitle{Massive Stars: Fund. Param. and Circumst. Interact.}

\title{Extragalactic Stellar Astronomy with Blue Supergiants} 

\author{
  N. Przybilla,\altaffilmark{1} 
  F. Bresolin,\altaffilmark{2}
  K. Butler,\altaffilmark{3}
  R.P. Kudritzki,\altaffilmark{2}
  M.A. Urbaneja\altaffilmark{2}
  and K.A. Venn\altaffilmark{4}}

\altaffiltext{1}{Dr. Remeis Sternwarte Bamberg, Sternwartstr. 7, 96049
Bamberg, Germany (przybilla@sternwarte.uni-erlangen.de).}

\altaffiltext{2}{Institute for Astronomy, 2680 Woodlawn Drive
Honolulu, Hawaii 96822, USA.}

\altaffiltext{3}{Universit\"ats-Sternwarte M\"unchen, Scheinerstr.~1, 
81679 M\"unchen, Germany.}

\altaffiltext{4}{University of Victoria,
Elliott Building, 3800 Finnerty Road, Victoria, BC, V8P 1A1, Canada}

\shortauthor{Przybilla et al.}
\shorttitle{Extragalactic Stellar Astronomy with Blue Supergiants}

\listofauthors{N. Przybilla, F. Bresolin, K. Butler, R.P. Kudritzki \& M.A.
Urbaneja}
\indexauthor{Przybilla, N.}
\indexauthor{Bresolin, F.}
\indexauthor{Butler, K.}
\indexauthor{Kudritzki, R.P.}
\indexauthor{Urbaneja, M.A.}
\indexauthor{Venn, K.A.}

\abstract{
The present generation of large telescopes facilitates spectroscopy of 
blue supergiants in galaxies out to distances beyond the Local Group. 
Recent developments in NLTE spectrum synthesis 
techniques allow for an accurate determination of stellar
parameters and chemical abundances.
Quantitative analyses of blue supergiants in different galactic environments
can provide tight observational constraints on:
{\sc i)} the evolution of massive stars over a wide range of
metallicities,
{\sc ii)} the chemical evolution of different galaxy types, using
stars as tracers of abundance gradients,
{\sc iii)} the extragalactic distance scale.
The current status of the field is summarised.
}

\resumen{
La generaci\'on actual de grandes telescopios permite el estudio
espectrosc\'opico de estrellas supergigantes azules en galaxias
incluso m\'as all\'a del Grupo Local. Recientes avancez en las
t\'ecnicas de s\'{\i}ntesis espectral en NLTE permiten una
determinaci\'on precisa de los par\'ametros estelares y las abundancias
qu\'{\i}micas. El an\'alisis cuantitativo de estas estrellas
en diversos entornos gal\'acticos puede proporcionar l\'{\i}mites
observacionales restrictivos en diversos campos: {\sc i)} la evoluci\'on
de estrellas masivas en un rango amplio de metalicicidades, {\sc ii)} la
evoluci\'on qu\'{\i}mica dediversos tipos de galaxias mediante
el uso de estas estrellas como trazadores de los gradientes de
abundancias, {\sc iii)} la escala de distancias extragal\'actica. En la
presente contribuci\'on resumimos el estado actual de este campo de
investigaci\'on.
}

\addkeyword{distance scale}
\addkeyword{galaxies: abundances}
\addkeyword{stars: abundances}
\addkeyword{stars: evolution}
\addkeyword{stars: fundamental parameters}
\addkeyword{supergiants}

\begin{document}
% Typeset article header
\maketitle

\section{Motivation}
Massive blue supergiants (BSGs) of spectral types B and A are among the visually
brightest stars in spiral and irregular galaxies. In the era of large
telescopes, this makes them primary candidates for spectroscopic studies 
even when situated in galaxies well beyond the Local Group. Information on 
fundamental stellar parameter and abundances for a wide variety of chemical
species (CNO, iron group elements, $\alpha$- and s-process elements) 
can thus be obtained. The challenges for the quantitative analysis are posed
by the high energy and momentum density of the radiation field in a 
tenuous atmosphere, requiring NLTE modelling techniques.

Analyses of samples of BSGs in different galactic environments
allow observational constraints on the evolution of massive stars and
galactochemical evolution to be derived. Abundances of the light elements
(He, CNO) act as tracers of rotational mixing (e.g. Maeder \& Meynet~2000) 
and may help to constrain the complex \mbox{(magneto-)}hydro\-dynamic processes 
relevant to stellar evolution empirically. The step to other galaxies is
required in order to test for metallicity effects. 
Furthermore, BSGs can act as tracers of abundance gradients, which allow us to 
discriminate between different models of galactochemical evolution (e.g.
Chiappini et al.~2001). The stellar results may be used to verify -- but
also to extend -- studies of H\,{\sc ii} regions. Finally, tighter
constraints on the extragalactic distance scale can be expected from
spectroscopic analyses of BSGs. Systematic uncertainties (metalli\-city,
reddening effects) that are of concern for classical photometric indicators
like the period-luminosity relationship for Cepheids may thus be avoided.

\section{Modelling \& Results}
We follow two approaches in modelling the atmospheres of BSGs. The
photospheric spectra of less-luminous A-type SGs can be modelled well
with classical line-blanketed LTE model atmospheres using NLTE line
formation for specific elements (Venn~1995). It has been shown that
this hybrid NLTE approach can be extended to objects of higher luminosity
and also into the regime of late B-type SGs (Przybilla et al.~2006). A new
generation of state-of-the-art NLTE model atoms (Przybilla \& Butler~2001,
2004; Przybilla et al.~2000, 2001ab; Nieva \& Przybilla~2006) allows stellar
parameters and elemental abundances to be derived with unprecedented
accuracy: effective temperature $T_{\rm eff}$ to better than 1--2\%,
surface gravity $\log g$ to 0.05--0.10\,dex and abundances to 
$\sim$0.05--0.10\,dex (random) and $\sim$0.10\,dex (systematic uncertainties).
Multiple NLTE ionization equilibria, Stark-broadened hydrogen profiles and
spectral energy distributions are simultaneously matched. Massive (NLTE)
spectrum synthesis allows us to reproduce practically the entire observed
high-resolution spectra in the visual, a prerequisite for applications of
the method to intermediate-resolution data, in that case however at reduced
accuracy. Early B-type SGs are modelled with
line-blanketed NLTE-model atmospheres accounting for hydrodynamical outflow
and spherical extension (Puls et al.~2005; Urbaneja et al.~2005b). This
allows the photospheric spectra and the stellar wind features to be analysed
in a self-consistent manner. Galactic BSGs are crucial test cases for the
analysis methodology, as detailed comparisons of theory with high-quality
spectra are feasible only here.  

Studies of abundance patterns of the light elements as tracers for rotational 
mixing in BSGs beyond the Milky Way have concentrated on the metal-poor SMC
(Venn~1999; Trundle \& Lennon~2005) and M\,33 (Urbaneja et al.~2005b). The
findings confirm predictions of stellar evolution calculations that the
efficiency of mixing processes increases with decreasing metallicity in a
qualitative way. However, there remain discrepancies between 
observational findings and predictions on the amount of mixing.

Abundance studies on heavier elements have been performed for a few objects in 
more distant galaxies of the Local Group at high spectral resolution using
Keck/HIRES and VLT/UVES: in M\,31, NGC\,6822 and WLM (Venn et al. 2000,
2001, 2003) and in Sextans\,A in the closeby Sextans-Antlia Group 
(Kaufer et al.~2004). Good agreement of stellar and nebular abundances
(mostly oxygen) is found in most cases, in both spiral and dwarf 
irregular galaxies. The low $[\alpha$/Fe$]$ ratios are consistent 
with the slow chemical evolution expected for dwarf galaxies.

Spectroscopic surveys at intermediate resolution using long-slit (WHT/ISIS) and 
multi-object spectrographs (VLT/FORS) have produced a considerable amount of data on
BSGs in the Local Group (M\,31: Trundle et al.~2002; M\,33: Urbaneja et al.~2005b; 
WLM: Bresolin et al.~2006). Beyond the Local Group, the investigations have concentrated so
far on the Sculptor Group spiral NGC\,300 (Bresolin et al.~2002; Urbaneja et
al.~2005a) and the field spiral NGC\,3621 (Bresolin et al.~2001) at a
distance of 6.6\,Mpc. The quantitative analysis of subsets of these stars finds
in general reasonably good agreement with published abundances from studies
of nebulae. However, systematic offsets may occur. In particular some statistical
indicators ($R_{23}$--O/H-calibrations) for nebular abundances are shown to
be prone to systematic error when compared to stellar abundances or nebular
abundances from H\,{\sc ii} regions with direct $T_{\rm e}$ determinations 
(Urbaneja et al.~2005a). Both, absolute abundances and the slope of
the abundance gradient may be affected.

Finally, it has been shown that BSGs can be used as spectroscopic 
distance indicators, via application of the wind momentum-luminosity
relationship (WLR, e.g. Kudritzki et al.~1999) and the flux-weighted
gravity-luminosity relationship (FGLR, Kudritzki et al.~2003).
Both methods have the potential to facilitate measurements of distance
moduli out to systems in the Virgo and Fornax clusters of galaxies with an
accuracy of 0.1\,mag or better, once properly calibrated. The intrinsic photometric
variability of BSGs have been shown to have a negligible effect on the FGLR
(Bresolin et al.~2004, 2006).

\end{document}